\documentclass[prl]{revtex4}

\usepackage{graphicx}
\usepackage{dcolumn}
\usepackage{bm}

\begin{document}
\renewcommand{\theequation}{\arabic{equation}}

\title{Spin and maximal acceleration\\}

\author{Giorgio Papini}
\altaffiliation[Electronic address:]{papini@uregina.ca}
\affiliation{Department of Physics and Prairie Particle Physics
Institute, University of Regina, Regina, Sask, S4S 0A2, Canada}


\begin{abstract}

We study the spin current tensor of a Dirac particle at accelerations close to the
upper limit introduced by Caianiello. Continual interchange between particle spin and angular momentum
is possible only when the acceleration is time-dependent. This represents a stringent
limit on the effect that maximal acceleration may have on spin physics in astrophysical applications.
We also investigate some dynamical consequences of
maximal acceleration.

\end{abstract}

 \maketitle

\setcounter{equation}{0}
\section{Introduction}

Recently, several aspects of spin physics have been actively
investigated. The interaction of photon \cite{ashby} and neutron spins \cite{demirel}
with non-inertial fields like rotation \cite{MASH1,MASH2,MASH3,PAP08} has been experimentally verified at the quantum level and
spin-induced effects for macroscopic objects have been observed \cite{Everitt,Iorio1,Iorio2}. It is also known that
spin is not a constant of motion when a particle interacts with external fields,
either electromagnetic \cite{grandy}, or gravitational \cite{PASP} and that continual
interchange between spin and orbital angular momentum is possible.

The purpose of this work is to study the behaviour of spin at large accelerations,
those that may be met close to a black hole and to the maximal acceleration (MA), an upper
limit introduced by Caianiello in his geometrical formulation of quantum mechanics \cite{cai1,cai2,cai3}.
In Caianiello's model, in fact, the absolute value of a particle proper acceleration satisfies
the inequality $a\leq \mathcal{A}_{m}$, where
$\mathcal{A}_{m}=2mc^{3}/\hbar$ is the upper limit mentioned and $m$ the
particle mass. No counterexamples are known
to the validity of this inequality. The limit $\hbar\rightarrow 0$ restores
$\mathcal{A}_{m}$ to its infinite, classical limit. The value of $\mathcal{A}_{m}$ is mass dependent
and very large even for the lightest particles. It leads to violations of the equivalence principle, also
a subject of great interest.

Classical and quantum arguments supporting the existence of a MA
have been given in the literature \cite{ca,pw,papinix,prove.1,prove.2,prove.3,prove.4,prove.5,prove.6,prove.7,prove.8,
prove.9,prove.10,prove.11,prove.11bis,prove.12,prove.13,prove.14,MASH2,prove.16,prove.17,prove.18,prove.19,prove.20,prove.21,prove.22,prove.23,prove.24,prove.25,prove.26,wh,b.1,b.2}.
MA is also found in the context of Weyl space \cite{pap.1,pap.2,pap.3,pap.4} and of a
geometrical analogue of Vigier's stochastic theory \cite{jv}. It rids
black hole entropy of ultraviolet divergences \cite{McG} and is a regularization procedure \cite{nesterenko}
that avoids the introduction of a fundamental length \cite{gs},
thus preserving the continuity of space-time.

A MA also exists in string theory \cite{gsv.1,gsv.2,gsv.3,gasp.1,gasp.2,fs,sa2} when the
acceleration induced by a background gravitational field reaches
the critical value $a_c = \lambda^{-1} = (\tilde{m}\alpha)^{-1}$ where
$\lambda$, $\tilde{m}$ and $\alpha^{-1}$ are string size, mass and
tension. At accelerations larger than $a_c$ the string extremities
become casually disconnected.

Applications of Caianiello's model include cosmology \cite{infl.1,infl.2,infl.3},
the dynamics of accelerated strings \cite{Feo.1,Feo.2}, neutrino
oscillations \cite{8,qua.1,qua.2} and the determination of a lower
neutrino mass bound \cite{neutrinobound}. The model also makes the
metric observer--dependent, as conjectured by Gibbons and Hawking
\cite{Haw}.

The model has been applied to classical \cite{sch} and quantum particles \cite{boson} falling in the
gravitational field of a collapsing, spherically symmetric object described
by the Schwarzschild metric
and also to the Reissner-Nordstr\"om \cite{reiss} and Kerr \cite{kerr} metrics.
In the model, the end product of stellar collapse is
represented by compact, impenetrable astrophysical objects whose
radiation characteristics are similar to those of known bursters
\cite{papiniz}.

The consequences of MA for the classical electrodynamics of
a particle \cite{cla}, the mass of the Higgs boson \cite{Higgs.1,Higgs.2},
the Lamb shift in hydrogenic atoms \cite{lamb}, muonic
atoms \cite{muo}, the
helicity and chirality of particles \cite{chen} and the tempertaure \cite{feol} have also been investigated.

Most recently Rovelli and Vidotto have found evidence for MA
and singularity resolution in covariant loop quantum gravity \cite{rovelli1},\cite{rovelli2}.

Caianiello's model is based on an embedding procedure \cite{sch}
that stipulates that the line element experienced by an
accelerating particle is represented by
\begin{equation} \label{eq1}
d\tau^2=\left(1+\frac{g_{\mu\nu}\ddot{x}^{\mu}\ddot{x}^{\nu}}{{\cal
A}_m^2}
\right)g_{\alpha\beta}dx^{\alpha}dx^{\beta}=\left(1-\frac{|a(x)|^{2}}{{\cal
A}_m^2}
 \right) ds^2\equiv
f(x) ds^2\,,
\end{equation}
where $g_{\alpha\beta}$ is a background gravitational field and $|a|$ the absolute value of the acceleration. The value $f=1$ corresponds to
the classical limit $\mathcal{A}\rightarrow \infty$ and $f=0$ to the MA limit.
A particle therefore experiences acceleration as if subjected to an external gravitational field
represented by the metric $g_{\mu\nu}=f(x)\eta_{\mu\nu}$, where $\eta_{\mu\nu}$ is the Minkowski metric (of signature -2), if the background is flat.
Particles of different mass experience different metrics, hence different effective gravitational fields, but their
kinematics is characterized by the same velocity field.
The metric (\ref{eq1}) lends support to geometrical models of confinement in the strong interactions and
hadronization processes. If an effective space-time curvature can be generated by acceleration, then confinement
inside hadrons can affect only quarks that are strongly accelerated by the strong interactions,
while other particles, leptons for instance, that are not affected by the strong interactions,
experience a geometry identical to that of an inertial observer.
Since Caianiello's model of quantum geometry
offers a metric to work with, it is convenient
to use it in conjunction with covariant wave equations that are the byproduct
of minimal coupling and Lorentz invariance.

Covariant wave equations that apply to particles with, or without spin,
have solutions \cite{PAP0,PAP1,PAP2,PAP3,PAP4} that are exact to first order
in the metric deviation $\gamma_{\mu\nu}= g_{\mu\nu}-\eta_{\mu\nu}$ and have been applied to problems like geometrical optics \cite{PAP3},
interferometry and gyroscopy \cite{PAP1}, the spin-flip of particles in gravitational and inertial fields \cite{PAP5},
radiative processes \cite{PAP6,PAP7} and spin currents \cite{PASP}.
We are interested in spin-1/2 particles, in particular. The covariant Dirac equation \cite{DE}
\begin{equation}\label{CDE}
  \left[i\gamma^{\mu}(x){\cal D}_\mu-m\right]\Psi(x)=0\,,
  \end{equation}
is remarkably successful in dealing
with all inertial and gravitational effects discussed in the literature \cite{COW,PW,BW,12, dinesh}.
The notations and units ($\hbar=c=1$) are as in \cite{PAP5}, in particular ${\cal D}_\mu=\nabla_\mu+i\Gamma_\mu (x)$, $\nabla_\mu$ is
the covariant derivative, $\Gamma_{\mu}(x)$ the spin connection, commas indicate partial derivatives
and the matrices $\gamma^{\mu}(x)$ satisfy the relations
$\{\gamma^\mu(x), \gamma^\nu(x)\}=2g^{\mu\nu}$. In the absence of external fields, (\ref{CDE}) reduces to the free Dirac equation
\begin{equation}\label{E}
\left(i\gamma^{\hat{\mu}}\partial_{\mu}-m\right)\psi_{0}(x)=0\,,
\end{equation}
where $\gamma^{\hat{\mu}}$ are the usual
constant Dirac matrices.

The first
order solution of (\ref{CDE}) is of the form
\begin{equation}\label{ED}
  \Psi(x) = {\hat T}(x) \psi_{0}(x)\,,
\end{equation}
where $\psi_{0}(x)$ is a solution of (\ref{E}), the operator $\hat{T}$ is given by
\begin{equation}\label{T}
    \hat{T}=
  -\frac{1}{2m}\left(-i\gamma^{\mu}(x)\mathcal{D}_{\mu}-m\right)e^{-i\Phi_{T}}\,,
\end{equation}
\begin{equation}\label{PHIS}
\Phi_{T}=\Phi_{S}+\Phi_{G},
\Phi_{S}(x)=\int_{P}^{x}dz^{\lambda}\Gamma_{\lambda}(z)\,,
\end{equation}
and
 \begin{eqnarray}\label{PH}
  \Phi_{G}(x) = -\frac{1}{4}\int_P^xdz^\lambda\left[\gamma_{\alpha\lambda,
  \beta}(z)-\gamma_{\beta\lambda, \alpha}(z)\right]\left[\left(x^{\alpha}-
  z^{\alpha}\right)k^{\beta}-\left(x^{\beta}-z^{\beta}\right)k^{\alpha}\right]+
 \frac{1}{2}\int_P^xdz^\lambda\gamma_{\alpha\lambda}(z)k^{\alpha}\,.
\end{eqnarray}
It is convenient to choose $\psi_{0}(x)$ in the form of plane waves, but wave packets can also be
used.

In (\ref{PHIS}) and (\ref{PH}), the path integrals are taken along
the classical world line of the fermion, starting from an arbitrary reference point
$P$ that will be dropped in the following. Only the path to $\mathcal{O}(\gamma_{\mu\nu})$ needs to be known in the integrations indicated
because (\ref{ED}) already is a first order solution. The positive energy solutions of (\ref{E}) are given by
\begin{equation}\label{psi0}
\psi(x)=u(\mathbf{k})e^{-ik_\alpha x^\alpha}=N
  \left(\begin{array}{c}
                \phi \\
                 \frac{\mathbf{\sigma}\mathbf{\cdot} \mathbf{k}}{E+m}\, \phi \end{array}\right)
                 \,e^{-ik_\alpha x^\alpha}\,,
\end{equation}
where $N=\sqrt{\frac{E+m}{2E}}$, $u^{+} u=1$, $\bar{u}=u^{+}\gamma^{0}$, $u^{+}_{1}u_{2}=u_{2}^{+}u_{1}=0$ and ${\bf \sigma}$ are the Pauli matrices.
In addition $\phi$ can take the forms $ \phi_{1}$  and $\phi_{2}$ where
$\phi_{1}=\left(\begin{array}{c}
1\\ 0 \end{array}\right)$, and $\phi_{2}=\left(\begin{array}{c}
0\\ 1 \end{array}\right)$.

Solution (\ref{CDE}) contains that of the covariant Klein-Gordon equation that, neglecting curvature dependent terms
becomes to $\mathcal{O}(\gamma_{\mu\nu})$
\begin{equation}\label{KG}
\left(\nabla_{\mu}\nabla^{\mu}+m^2\right)\phi(x)\simeq\left[\eta_{\mu\nu}\partial^{\mu}\partial^{\nu}+m^2
+\gamma_{\mu\nu}\partial^{\mu}\partial^{\nu}
\right]\phi(x)- \frac{1}{2}\eta^{\sigma\rho}\left(2\gamma_{\rho,\mu}^{\mu}-\gamma_{,\rho}\right)\phi=0\,,
\end{equation}
where $\gamma\equiv \gamma_{\rho}^{\rho}$.
The solution of (\ref{KG}) is obtained by solving the Volterra
equation
\begin{equation}\label{V}
\phi(x)=\phi_{0}(x)-\int_P^{x}d^4x' G(x,x') \gamma_{\mu\nu}(x')
\partial^{'\mu}\partial^{'\nu}\phi(x')\,,
\end{equation}
along the particle world-line, where $P$ is again a fixed reference point, $x$ a generic point in
the physical future along the world-line, $G(x,x') $ is the causal Green function with
$(\partial^2 +m^2)G(x,x')=\delta^4 (x-x') $. The free Klein-Gordon equation is
\begin{equation}\label{KG0}
(\partial^2 +m^2)\phi_{0}=0\,.
\end{equation}
In first approximation
$\phi_{0}$ can be substituted for $\phi$ in (\ref{V}) and the integrations can then be carried following \cite{PAP1,PAP2}.
The solution of (\ref{KG}) is
\begin{equation}\label{SO}
\phi(x)=\left(1-i\Phi_{G}(x)\right)\phi_{0}(x)\,,
\end{equation}
which is contained in $\exp(-i\Phi_{T})$. Higher order approximations to the solution of (\ref{SO}), therefore of
(\ref{CDE}), can be obtained by writing
\begin{equation}\label{IT}
\phi(x)=\Sigma_{n}\phi_{(n)}(x)=\sum_{n}e^{-i\Phi_{G}}\phi_{(n-1)}(x)\,.
\end{equation}
Because of the structure of (\ref{IT}), the higher order corrections are expected to be well behaved and to not affect the conclusions.

\section{Spin currents}

The transfer of angular momentum between the external field and the fermion spins can be calculated using the
spin current tensor \cite{grandy}
\begin{equation}\label{ST}
S^{\rho\mu\nu}=\frac{1}{4im}\left[\left(\nabla^{\rho}\bar{\Psi}\right)\sigma^{\mu\nu}(x)\Psi-\bar{\Psi}\sigma^{\mu\nu}(x)\left(\nabla^{\rho}\Psi\right)\right]\,,
\end{equation}
that satisfies the conservation law $S^{\rho\mu\nu},_{\rho}=0 $ when all $\gamma_{\alpha\beta}(x)$ vanish and yields in addition the expected result $S^{\rho\mu\nu}= \frac{1}{2}\bar{u}_{0}\sigma^{\hat{\mu}\hat{\nu}}u_{0}$ in the rest frame of the particle. Writing $\sigma^{\mu\nu}(x)\approx \sigma^{\hat{\mu}\hat{\nu}}+h^{\mu}_{\hat{\tau}}\sigma^{\hat{\tau}\hat{\nu}}+h^{\nu}_{\hat{\tau}}\sigma^{\hat{\mu}\hat{\tau}}$,
where $\sigma^{{\hat \alpha}{\hat \beta}}=\frac{i}{2}[\gamma^{\hat
\alpha}, \gamma^{\hat \beta}]$, substituting (\ref{E}) and (\ref{T}) in (\ref{ST})
one obtains, to $\mathcal{O}(\gamma_{\alpha\beta})$,
\begin{equation}\label{STF}
S^{\rho\mu\nu}=\frac{1}{16im^3}\bar{u}_{0}\left\{8im^2k^{\rho}\sigma^{\hat{\mu}\hat{\nu}}+8imk^{\rho}h^{[\mu}_{\hat{\tau}}\sigma^{\hat{\tau}\hat{\nu}]}+ \right.
\end{equation}
\[
\left.
4imk^{\rho}\left(\Phi_{G,\alpha}
+k_{\sigma}h^{\sigma}_{\hat{\alpha}}\right)\left\{\sigma^{\hat{\mu}\hat{\nu}},\gamma^{\hat{\alpha}}\right\}-8imk^{\rho}\Phi_{G}k^{[\mu}\gamma^{\hat{\nu}]}+ \right.
\]
\[
\left.
4mk^{\rho}k_{\alpha}\left[\sigma^{\hat{\mu}\hat{\nu}},\left(\gamma^{\hat{\alpha}}\Phi_{S}-\gamma^{\hat{0}}\Phi^{+}_{S}
\gamma^{\hat{0}}\gamma^{\hat{\alpha}}\right)\right]+
4m^2k^{\rho}\left[\sigma^{\hat{\mu}\hat{\nu}},\left(\Phi_{S}-\gamma^{\hat{0}}\Phi^{+}_{S}\gamma^{\hat{0}}\right)\right]- \right.
\]
\[
\left. 8m^2 k^{\rho}h^{0}_{\hat{\alpha}}\left[\gamma^{\hat{0}},\left[\sigma^{\hat{0}\hat{\alpha}},\sigma^{\hat{\mu}\hat{\nu}}\right]\right] -8im^2k_{\sigma}\left(\Gamma^{\sigma}_{\alpha\beta}\eta^{\beta\rho}+\partial^{\rho}h^{\sigma}_{\hat{\alpha}}\right)\eta^{\alpha[\mu}\gamma^{\hat{\nu}]}+ \right.
\]
\[
\left. 8im^2\partial^{\rho}\Phi_{G}\left(4m\sigma^{\hat{\mu}\hat{\nu}}-2ik^{[\mu}\gamma^{\hat{\nu}]}\right)+4im^2\gamma^{\hat{0}}\Gamma^{\rho+}\gamma^{\hat{0}}\left\{\left(\gamma^{\hat{\alpha}}
k_{\alpha}+m\right),\sigma^{\hat{\mu}\hat{\nu}}\right\}\Gamma^{\rho} \right\}u_{0}
\]
where use has been made of the relation
\begin{equation}\label{GMUNU}
 \Phi_{G,\mu\nu}=k_{\alpha}\Gamma^{\alpha}_{\mu\nu}\,.
 \end{equation}
It is therefore possible to separate $S^{\rho\mu\nu}$ in inertial and non-inertial parts. The first term on the r.h.s. of (\ref{STF}) gives the usual result in the particle rest frame, when the external field vanishes.
From (\ref{eq1}) we get
\begin{equation}\label{mde}
\gamma_{\mu\nu}=(f(x)-1)\eta_{\mu\nu}\,.
\end{equation}
To first order the tetrad is given by
\begin{eqnarray} \label{tet}
e^{\mu}_{\hat{\alpha}}&\approx &\delta^{\mu}_{\alpha}+h^{\mu}_{\hat{\alpha}}\,\,, h_{\hat{\alpha}}^{\nu}=\delta_{\alpha}^{\nu}\left(\frac{1}{\sqrt{f}}-1\right)\,,
\end{eqnarray}
from which the spinorial connection can be calculated using the relations
\begin{equation}\label{II.2}
  \gamma^\mu(x)=e^\mu_{\hat \alpha}(x) \gamma^{\hat
  \alpha}\,,\qquad
  \Gamma_\mu(x)=-\frac{1}{4} \sigma^{{\hat \alpha}{\hat \beta}}
  e^\nu_{\hat \alpha}\nabla_{\mu}e_{\nu\hat{\beta}}\,.
\end{equation}
The result is
\begin{equation}\label{GA}
\Gamma_{\mu}=\sigma^{\hat{\alpha}\hat{\beta}}\eta_{\alpha\mu}\left(\frac{1}{2}\ln f\right)_{,\beta}\,.
\end{equation}
The choice $\phi =\phi_{1}$ corresponds to $u_1$ and  $\phi =\phi_2$ to $u_2$. Substituting in (\ref{psi0}), one finds
\[u_1=N
 \left(\begin{array}{c}
                1\\
                0\\
                \frac{k^3}{E+m}\\
                \frac{k^1 +ik^2}{E+m}\end{array}\right)\,,
                u_2=N
\left(\begin{array}{c}
                0\\
                1\\
                \frac{k^1-ik^2}{E+m}\\
                \frac{-k^3}{E+m}\end{array}\right)\,,\]
that are not eigenspinors of the matrix $\Sigma^3 =\sigma^3 I$ whose eigenvalues represent the spin components in
the $z$-direction, but become eigenspinors of $\Sigma^3$ when $k^1 =k^2 =0$, or in the rest frame of the fermion $\textbf{k}=0$.
By performing a transformation of coordinates
$x_{\mu}\rightarrow x_{\mu}+\xi_{\mu}$, where $\xi_{\mu}$ is small of first order,
we obtain $\gamma_{\mu\nu}\rightarrow \gamma_{\mu\nu}-\xi_{\mu,\nu}-\xi_{\nu,\mu}$ and write the Lanczos-DeDonder condition
in the form
\begin{equation}\label{LDD}
\gamma_{\alpha,\nu}^{\nu}-\frac{1}{2}\gamma_{,\alpha}\rightarrow \gamma_{\alpha,\nu}^{\nu}-\frac{1}{2}\gamma_{,\alpha}-\partial_{\nu}\partial^{\nu}\xi_{\alpha}-f_{,\alpha}=0\,.
\end{equation}
By choosing $\xi_{\alpha}$ to satify $\partial_{\beta}\partial^{\beta}\xi_{\alpha}+f_{,\alpha}=0$,
we get $\partial^{\mu}\Phi_{G,\mu}=k_{\alpha}\Gamma_{\mu\nu}^{\alpha}\eta^{\mu\nu}=0$. We also find
$\partial_{\alpha}\partial^{\alpha}\Phi_{G,\beta}=k_{\beta}\partial_{\alpha}\partial^{\alpha}f /2\,, \Gamma_{\alpha\varrho}^{\beta}=\eta^{\beta\sigma}(\eta_{\sigma\alpha}f_{,\varrho}+
\eta_{\sigma\varrho}f_{,\alpha}-\eta_{\alpha\varrho}f_{,\sigma})/2$, and the spinorial connection gives $\partial^{\rho}\Gamma_{\rho}=0$ except
at $f=0$ for which the external field approximation breaks down.
With these simplifications, setting $k_{1}=k_{2}=k_{3}=0$ and by differentiating (\ref{ST}) with respect to $x_{\rho}$, we finally find
\begin{equation}\label{CON}
\partial_{\rho}S^{\rho\mu\nu}= \frac{\bar{u}_{0}}{16im^{3}}\left[8im^3\partial_{0}B^{\mu\nu}+4im^2\left(\Phi_{G,\alpha0}+m\partial_{0}h^{0}_{\hat{\alpha}}\right)
\left\{\sigma^{\hat{\mu}\hat{\nu}},\gamma^{\hat{\alpha}}\right\} \right.
\end{equation}
\[\left.
-8im^2\left(k^{\mu}\gamma^{\hat{\nu}}-k^{\nu}\gamma^{\hat{\mu}}\right)
+4m^3\left(\sigma^{\hat{\mu}\hat{\nu}}\gamma^{\hat{0}}\Gamma_{0}-
\gamma^{\hat{0}}\Gamma_{0}^{\dag}\sigma^{\hat{\mu}\hat{\nu}}\right)\right.
\]
\[\left.
+4m^3 \left(\sigma^{\hat{\mu}\hat{\nu}}\Gamma_{0}-\gamma^{\hat{0}}\Gamma_{0}^{\dag}\gamma^{\hat{0}}\sigma^{\hat{\mu}\hat{\nu}}\right)+
4im^3\partial_{0}h_{\hat{\alpha}}^{0}\left(\gamma^{\hat{0}}\gamma^{\hat{\alpha}}\sigma^{\hat{\mu}\hat{\nu}}-\sigma^{\hat{\mu}\hat{\nu}}\gamma^{\hat{\alpha}}\gamma^{\hat{0}}\right)\right.
\]
\[\left.
-4im\partial_{\rho}\partial^{\rho}\Phi_{G,\alpha}\left(\eta^{\alpha\mu}\gamma^{\hat{\nu}}-\eta^{\alpha\nu}\gamma^{\hat{\mu}}\right)
-8im^3\left(\sigma^{\hat{\mu}\hat{\nu}}\Gamma_{0\rho}^{\rho}+\sigma^{\hat{\alpha}\hat{\nu}}\Gamma_{\alpha0}^{\mu}+\sigma^{\hat{\mu}\hat{\alpha}}\Gamma_{\alpha0}^{\nu}\right)
\right] u_{0}\,,
\]
where $B^{\mu\nu}\equiv h^{\mu}_{\hat{\tau}}\sigma^{\hat{\tau}\hat{\nu}}+h_{\hat{\tau}}^{\nu}\sigma^{\hat{\mu}\hat{\tau}}$.
The only non-vanishing components of (\ref{CON}) are
\begin{equation}\label{COMP}
\partial_{\rho}S^{\rho 12}=-\frac{3f_{,0}}{4f^{\frac{3}{2}}}-f_{,0}\simeq-\frac{9}{8}f_{,0}\,\,,\partial_{\rho}S^{\rho01}\simeq \frac{1}{4}f_{,2}\,\,,
\partial_{\rho}S^{\rho02}\simeq -\frac{1}{4}f_{,1}\,.
\end{equation}
The derivatives of $f=1-(|a|/\mathcal{A})^2$, rather than $f$ itself, are responsible for the interchange of spin and angular momentum.
In fact, no interchange is possible for a strictly uniform acceleration \cite{BINI}.
The interchange can take place for any value of the acceleration for $0< f < 1$.
We notice that in order to alter the component $S^{\rho}_{12}$, that in the rest system of the particle corresponds
to the spin density in the direction of motion, simple flow of momentum between field and
particle is not sufficient. A time-dependent acceleration is necessary, and that requires that energy be transferred
from the external accelerating agent. It is therefore not so much the acceleration that affects
the spin-angular momentum interchange, as the way acceleration is applied.
The remaining two components of $\partial_{\rho}S^{\rho\mu\nu}$ refer to the motion of the particle as a whole
and do not require any time dependence of $f$. These results do not depend on any specific model for $f$.

\section{Dispersion relations and particle motion}

Some considerations about particle motion in the MA model are now in order. We are not concerned with spin in this section and consider
a spinless, uncharged particle for simplicity.
By using Schroedinger's logarithmic transformation $\phi = e^{-iS} $ \cite{LANC}, we can pass from the KG equation (\ref{KG}) to the quantum Hamilton-Jacobi
equation. We find to $O(\gamma_{\mu\nu})$
\begin{equation}\label{QHJ}
i(\eta^{\mu\nu}-\gamma^{\mu\nu})\partial_{\mu}\partial_{\nu}S-(\eta^{\mu\nu}-\gamma^{\mu\nu})\partial_{\mu}S\partial_{\nu}S +m^2=0 \,,
\end{equation}
where
\begin{equation}\label{S}
S= k^{\beta}\left\{x_{\beta}+\frac{1}{2}\int^{x}dz^{\lambda}\gamma_{\beta\lambda}(z)-\frac{1}{2}\int^{x}dz^{\lambda}\left(\gamma_{\alpha\lambda,\beta}(z)-\gamma_{\beta\lambda,\alpha}(z)\right)
\left(x^{\alpha}-z^{\alpha}\right)\right\}\,.
\end{equation}
It is known that the Hamilton-Jacobi equation is equivalent to Fresnel's wave equation in the limit of large
frequencies \cite{LANC}. However, at smaller, or moderate frequencies the complete  equation (\ref{QHJ}) should be used. We follow
this path. By substituting (\ref{S}) into the first term of (\ref{QHJ}), we obtain
\begin{equation}\label{ldd}
i(\eta^{\mu\nu}-\gamma^{\mu\nu})\partial_{\mu}\partial_{\nu}S=i\eta^{\mu\nu}\partial_{\mu}(k_{\nu}+\Phi_{G,\nu})-i\gamma^{\mu\nu}\partial_{\mu}k_{\nu}=i\eta^{\mu\nu}\Phi_{G,\mu\nu}=i
k_{\alpha}\eta^{\mu\nu}\Gamma^{\alpha}_{\mu\nu}=0 \,,
\end{equation}
on account of (\ref{LDD}). This part of (\ref{QHJ}) is usually neglected in the limit $\hbar\rightarrow 0$.
Here it vanishes as a consequence of (\ref{S}). The remaining terms of (\ref{QHJ}) yield the classical Hamilton-Jacobi equation
\begin{equation}\label{CHJ}
(\eta^{\mu\nu}-\gamma^{\mu\nu})\partial_{\mu}S\partial_{\nu}S -m^2=\gamma^{\mu\nu}k_{\mu}k_{\nu}-2k^{\mu}\Phi_{G,\mu}=0 \,,
\end{equation}
because $k^{\mu}\Phi_{G,\mu}=1/2 \gamma^{\mu\nu}k_{\mu}k_{\nu}$. Equation (\ref{SO}) is  therefore a solution of the more general quantum
equation (\ref{QHJ}). It also follows that the particle acquires a generalized "momentum"
\begin{equation}\label{mom}
P_{\mu}=k_{\mu}+\Phi_{G,\mu}=k_{\mu}+\frac{1}{2}\gamma_{\alpha\mu}k^{\alpha}-\frac{1}{2}\int^{x}dz^{\lambda}\left(\gamma_{\mu\lambda,\beta}(z)-
\gamma_{\beta\lambda,\mu}(z)\right)k^{\beta}\,,
\end{equation}
that describes the geometrical optics of particles correctly and gives the correct deflection predicted by general relativity.
It is Feynman's "p-momentum" in the case of gravity and gravity-like fields.
On using the relation
$\Phi_{G,\mu\nu}=k_{\alpha}\Gamma_{\mu\nu}^{\alpha}$
and differentiating (\ref{mom}) we obtain the covariant derivative of $P_{\mu}$
\begin{equation}\label{acc}
\frac{DP_{\mu}}{Ds}=m\left[\frac{du_{\mu}}{ds}+\frac{1}{2}\left(\gamma_{\alpha\mu,\nu}-\gamma_{\mu\nu,\alpha}+\gamma_{\alpha\nu,\mu}\right)u^{\alpha}u^{\nu}\right]
\end{equation}
\[=m\left(\frac{du_{\mu}}{ds}+\Gamma_{\alpha,\mu\nu}u^{\alpha}u^{\nu}\right)=
\frac{Dk^{\mu}}{Ds}\,.\]
This result is independent of any choice of field equations for $\gamma_{\mu\nu}$. We see from (\ref{acc}) that if $k_{\mu}$ follows a geodesic, then $DP_{\mu}/Ds=0$  and  $D(P_{\alpha}P^{\alpha})/Ds=0$.
The classical equations of motion are therefore contained in (\ref{acc}), but it would require the particle described by (\ref{KG}) to just choose a geodesic, among all
paths allowed to a quantum particle. It also follows from (\ref{mom}) that
\begin{equation}\label{mom1}
\frac{D(P_{\alpha}P^{\alpha})}{Ds}=2m^2 \frac{df}{ds}\,,
\end{equation}
and that, therefore, $P_{\alpha}D(P^{\alpha}/Ds)\neq 0$. For massless bosons, however, $P_{\alpha}$ and $DP_{\alpha}/Ds$ are still orthogonal.
Remarkably, (\ref{mom}) is an exact integral of (\ref{acc})
which can itself be integrated to give the particle's motion
\begin{equation}\label{U}
X_{\mu}=x_{\mu}+\frac{1}{2}\int^{x}dz^{\lambda}\left\{\gamma_{\mu\lambda}-\left(\gamma_{\alpha\lambda,\mu}-\gamma_{\mu\lambda,\alpha}\right)\left(x^{\alpha}-z^{\alpha}\right)\right\}\,.
\end{equation}

By substituting the explicit espressions for $\gamma_{\mu\nu}$ in (\ref{mom}) and (\ref{U}), we obtain
\begin{equation}\label{PX}
P_{\mu}=k_{\mu}+\frac{m}{2}\int^{s}ds f_{,\mu}\,\,\,, X_{\mu}=x_{\mu}+\frac{1}{2m}\int^{s}ds \left\{k_{\mu}\left(f-1\right)-k_{\alpha}\left(x^{\alpha}-z^{\alpha}\right)f_{,\mu}+
k_{\mu}f_{,\alpha}\left(x^{\alpha}-z^{\alpha}\right)\right\}\,.
\end{equation}
There is no back reaction to the motion described by (\ref{U}) and (\ref{mom}). The field experienced by the particle
is its own acceleration and that takes into account any back reaction automatically.
This can also be seen as follows.
The back reaction is normally introduced by means of a four-vector \cite{LL}
$g_{\mu}$ such that $g_{\mu}P^{\mu}=0$.
A natural choice for $g_{\mu}$ is
\begin{equation}\label{GMU}
  g_{\mu}=m^2 \frac{D^{2}P_{\mu}}{Ds^2}-P_{\mu}P_{\alpha}\frac{D^{2} P\alpha}{Ds^2}\approx m^2 \frac{D^{2}P_{\mu}}{Ds^2}-k_{\mu}k_{\alpha}\frac{D^{2}P^{\alpha}}{Ds^2}
\end{equation}
to $O(\gamma_{\alpha\beta})$. We obtain, in fact,
\begin{equation}\label{GMU1}
g_{\mu}=m^2 \left[\Phi_{G,\mu\nu\alpha}u^{\alpha}u^{\nu}+\Phi_{G,\mu\nu}\frac{du^{\nu}}{ds}\right]-k_{\mu}k^{\alpha}\left[\Phi_{G,\alpha\nu\sigma}u^{\sigma}u^{\nu}+
\Phi_{G,\alpha\nu}\frac{du^{\nu}}{ds}\right]\,,
\end{equation}
from which $g_{\mu}P^{\mu}=0$ follows. By substituting (\ref{mde}) in (\ref{GMU1}), we obtain
\begin{equation}\label{gmu}
g_{\mu}=\frac{m^{2}}{2}\left(\frac{df_{,\mu}}{ds}-k_{\mu}\frac{d^{2}f}{ds^{2}}\right)\,.
\end{equation}
which along a particle world-line gives $g_{\mu}k^{\mu}/m=0$ and also $\int g_{\mu}dx^{\mu}=0$, as expected, because no energy-momentum dissipation mechanism
is provided. The situation would of course change if the particle were charged.

\section{Conclusions}

We have investigated a particle spin at accelerations
close to the MA limit. The model incorporates MA
by assuming that particles are subjected to a metrical field
that has a gravity-like behaviour that changes from particle to particle according
to a particle's mass. The model also provides a comprehensive framework to treat accelerations
with values between the classical and the MA limits.

The solutions of the covariant Dirac equation,
consisting of plane wave solutions of the free Dirac equation and appropriate spin terms, have been applied to
the third rank spin current tensor. The calculations are performed in the rest frame of the fermion
and confirm that continual interchange between spin and angular momentum occurs in the case of MA fields,
but only if the acceleration is time-dependent ($f_{,0}\neq 0$). This requires that a transfer of energy from the
agent responsible for the acceleration, say a very compact star, or a black hole and the particle
take place because the field is not stationary.
The non vanishing components $\partial_{\rho}S^{\rho01}$ and $\partial_{\rho}S^{\rho02}$
of the spin current tensor refer to the motion of the particle as a whole and are present whatever the nature of the acceleration.
Even in the case of MA, uniform acceleration produces no observable effects on the particle spin, in agreement with \cite{BINI}.
These results are independent of any model for $f$ and provide a stringent limit on the the
astrophysical applications of spin physics, even in the case of MA.

The back reaction of the MA field on the particle vanishes along the particle world-line
and momentum cannot therefore be dissipated unless the particle is charged, or another specific radiation mechanism is provided.

\end{document}